\documentstyle[epsfig]{aipproc}

\begin{document}
\title{X-ray Spectroscopy of Bursts from SGR1806-20 with RXTE}

\author{Tod E. Strohmayer$^1$, and Alaa Ibrahim$^2$}
\address{$^1$LHEA, NASA/GSFC, Greenbelt, MD 20771\thanks{USRA research scientist}\\
$^2$Dept. of Physics, University of Maryland, College Park, MD}
\maketitle
\begin{abstract}

We report on new {\it Rossi X-ray Timing Explorer} (RXTE) X-ray spectral
analysis of bursts from SGR1806-20, the most prolific SGR source known. 
Previous studies of bursts from this source noted the remarkable lack of 
spectral variability both in single bursts as well as from burst to burst.
Although we find that the spectrum both within and among bursts is quite uniform
we do find evidence for significant spectral changes within bursts as well as
from burst to burst. We find that optically thin thermal bremsstrahlung spectra
(OTTB) including photoelectric absorption provide the best fits to most bursts, however, other models (power law, Band GRB model) can also produce 
statistically acceptable fits. We confirm the existence of a rolloff in the
photon number spectrum below 5 keV. When modelled as photoelectric absorption
and OTTB the inferred column is between $0.8 - 1.2 \times 10^{23}$ cm$^{-2}$.
This value is larger than the $\approx 0.6 \times 10^{23}$ cm$^{-2}$ inferred
from ASCA observations of the persistent X-ray counterpart, but less than the
$\approx 10.0 \times 10^{23}$ cm$^{-2}$ indicated by ICE data. 

\end{abstract}

\section*{Introduction}

Soft Gamma-Ray Repeaters (SGR) are a rare class of recurrent high energy
transient believed to be associated with young neutron stars \cite{KF,Mur}. SGR1806-20 is the
most prolific object in this class. Between 1978 and 1986 more than one hundred
burst events were detected from this source by X-ray detectors on the
International Cometary Explorer (ICE) satellite \cite{FLU94,Ulmer}. Various
bursts from this source have also been detected by many of the interplanetary
network instruments \cite{Norris}. BATSE detected a reactivation of SGR1806-20
in November, 1996 \cite{Kouv96a}. This triggered pointed observations of this source with RXTE in an effort to investigate burst spectra, search for pulsations, extend the burst size distribution to much fainter levels and
investigate the spectrum
of the persistent emission. We obtained a set of 4 pointed observations totaling
100 ksec of on source time. Observations from Nov. 18, 03:25:00 UTC to 11:30:00
UTC revealed more than 100 SGR bursts ranging in peak flux from 
$\approx 2 \times 10^{-9}$ ergs cm$^{-2}$ s$^{-1}$ to $>$ $1 \times 10^{-6}$
ergs cm$^{-2}$ s$^{-1}$. Here we summarize some of the X-ray spectral properties
of these bursts as measured with the proportional counter array (PCA).

\section*{Spectral Analysis}

A striking feature of SGR bursts noted from previous studies is the uniformity
of spectra from burst to burst and within a single event
\cite{Kouv,FLU94,Norris}.
With its large area, low background and high time resolution PCA observations
can test the uniformity of SGR burst spectra to a much greater level than
previous instruments. 

\begin{figure}[b!] % fig 1
\centerline{\epsfig{file=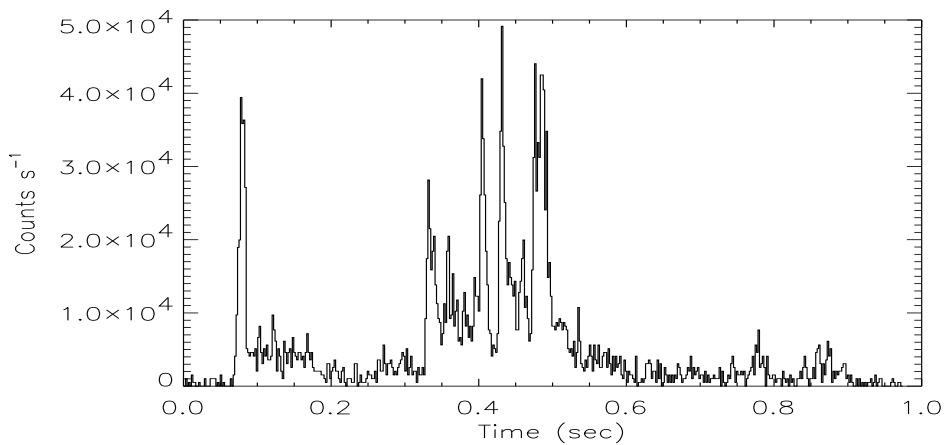,height=2.5in,width=5.0in}}
\vspace{10pt}
\caption{Time history of a typical burst from SGR1806-20 as seen in the PCA. The
peak countrate is near 50,000 cts/sec. Note that the typical PCA background
rate is 120 cts/sec.}
\label{fig1}
\end{figure}

We have not yet completed an exhaustive investigation of burst
spectra, however, we have investigated a sample of moderately bright bursts for
which the PCA deadtime is not excessive. We give examples of spectral variability from burst to burst as well as evidence for significant spectral
evolution in a single event. 
We obtained X-ray event data with 125 $\mu$s (1/8192 sec)
time resolution and 64 spectral channels across the 2 - 100 keV PCA bandpass.
For each burst we estimated the background using about 20 s of pre- and
post-burst data. In all cases the backgrounds were flat. For the bursts described here the background was at most a few percent of
the total counts. Figure 1 shows the time history of one of the bursts detected
in the PCA.
 
\begin{figure}[b!] % fig 2
\centerline{\epsfig{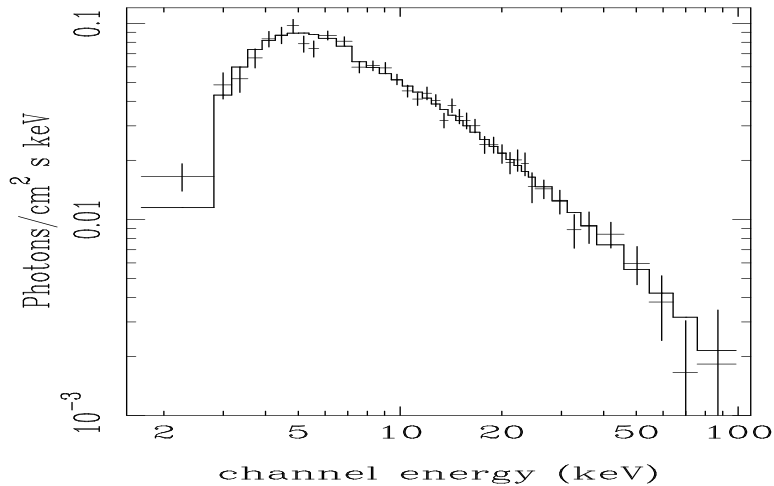}}
\vspace{10pt}
\caption{Best fit spectral model for the burst shown in Figure 1. The model includes thermal bremsstrahlung (OTTB) modified by photoelectric absorption. The
best fit parameters and confidence region for this burst are shown in figure 3.}
\label{fig2}
\end{figure}

We fit spectra with optically thin thermal bremsstrahlung (OTTB) and power law
models both modified by photelectric absorption. Both models provide 
statistically acceptable
fits to most bursts, but the OTTB model provides a marginally better fit in
almost all cases. We also fit the gamma-ray burst (GRB)
model which has now been used extensively to investigate the continuum spectra
of GRB with BATSE \cite{B93}, but the OTTB model also provides a marginally better fit than this model. Figure 2 shows the best fit OTTB model for the burst
in figure 1. The best fit temperature, $kT$, and column density of Hydrogen, $n_H$, for this burst are 164 keV and $12 \times 10^{22}$ cm$^{-2}$. We tested
for spectral variability from burst to burst by comparing the derived confidence
regions for the OTTB model parameters for different bursts. As we show below,
different bursts do show statistically significant differences in the derived
OTTB model parameters. To illustrate this we compare in Figure 3 the derived
confidence regions for two bursts which had similar peak countrates and
durations. We compare bursts with similar peak countrates in order to reduce any
spectral changes that could be introduced by differential deadtime effects.
Even at peak rates of 50,000 cts/sec the deadtime fraction is not more than
about 18 \%, and is reasonably well understood \cite{Jah}. We show the 68, 90,
and 99 \% confidence contours for each burst. The contours were computed by
calculating the $\delta\chi^2$ appropriate for three parameters ($kT$, $n_H$, and normalization constant) \cite{LMB76}. The contours centered at $kT = 164$
keV and $n_H = 12 \times 10^{22}$ cm$^{-2}$ are those for the burst shown in
figures 1 and 2. The confidence regions for these two bursts are disjoint at the
99.9 \% level, suggesting that these two bursts had measurably different
spectra. This result is not unique to these two bursts, across the sample of
bursts analysed to date we find that values for $kT$ and $n_H$ can range from
$\approx 20 - 120$ keV to $\approx 7 - 13 \times 10^{22}$ cm$^{-2}$,
respectively.
 
\begin{figure}[b!] % fig 3
\centerline{\epsfig{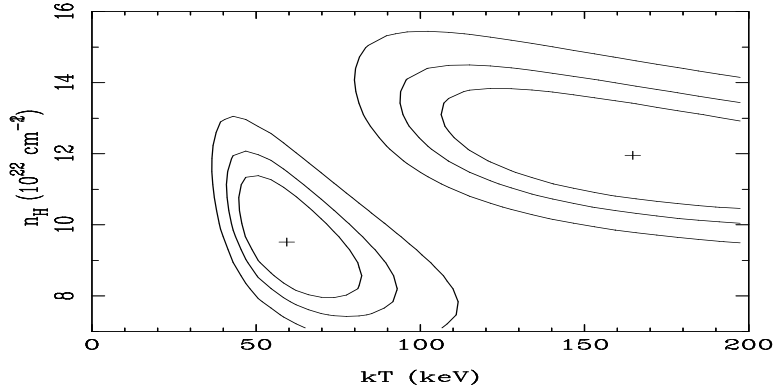}}
\vspace{10pt}
\caption{Confidence regions for $kT$ and $n_H$ from the OTTB model for two 
different bursts. The contours centered at $kT = 164$ keV and $n_H = 12$ are
those for the burst shown in figure 1.}
\label{fig3}
\end{figure}

\begin{figure}[b!] % fig 4
\centerline{\epsfig{file=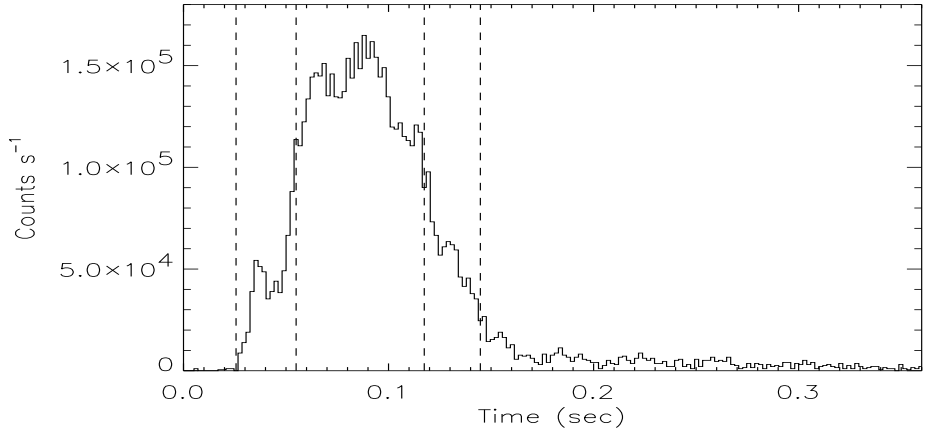,height=2.5in,width=5.0in}}
\vspace{10pt}
\caption{Time history of the burst for which we compared spectra during the 
rise and fall. The vertical dashed lines denote the intervals over which the 
spectra were accumulated.}
\label{fig4}
\end{figure}

\begin{figure}[b!] % fig 5
\centerline{\epsfig{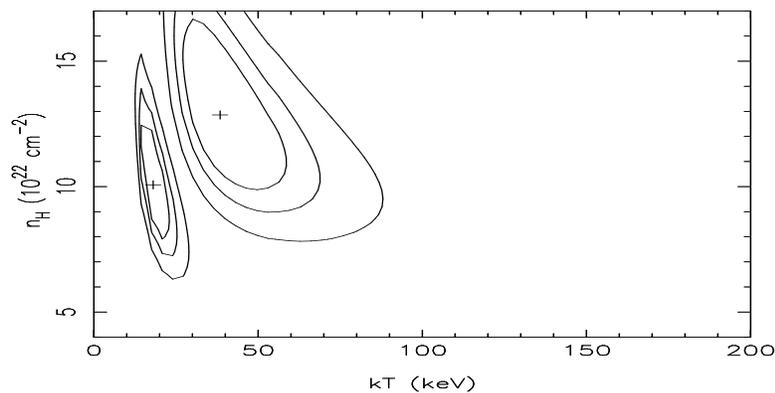}}
\vspace{10pt}
\caption{Confidence regions for the OTTB model fits to the rising and falling
intervals for the burst in figure 4. The contours denote the 68, 90, and 99 \%
confidence regions. The rising interval contours have the higher $kT$.}
\label{fig5}
\end{figure}

We have also investigated spectral variations within bursts. Figure 4 shows the
time history of a bright burst which shows evidence for spectral evolution.
We accumulated spectra both at the rising and falling edges of this burst. The
accumulation intervals are denoted by the vertical dashed lines in figure 4.
We selected the intervals to have approximately the same total number of counts
as well as similar countrate profiles to again minimize any differential
deadtime effects. Figure 5 compares the confidence regions for the OTTB model
parameters for both the rising and falling portions of the burst.  The rising 
interval is represented by the contours with the higher (harder) temperature.
The regions are disjoint at the $\approx 99.5 \%$ level, suggesting that the 
spectrum during the rising portion of this burst was moderately harder than 
during the falling portion. To our knowledge this is the first substantial 
evidence for spectral variations during bursts from SGR1806-20.

%\subsection*{The low energy rollover}

Spectral results based on ICE data from SGR1806-20 showed strong evidence for a
rolloff in the burst spectra below about 8 - 10 keV \cite{FLU94}. Our results
confirm the presence of a downturn in the photon spectrum at about 4 - 5 keV.
All models we investigated required with high significance such a rolloff in the
photon number spectrum. Our modelling with a photelectric absorption component
implies an absorbing column of $8 - 12 \times 10^{22}$ cm$^{-2}$. This is higher
than the $\approx 6 \times 10^{22}$ cm$^{-2}$ inferred from ASCA measurements
\cite{Mur}, but significantly less than the column of about $100.0 \times
10^{22}$ cm$^{-2}$ suggested from the analysis of ICE data \cite{FLU94}.

\end{document}